\documentstyle[12pt]{article}
\begin{document}
\def\beq{\begin{equation}}
\def\eeq{\end{equation}}
\def\beqa{\begin{eqnarray}}
\def\eeqa{\end{eqnarray}}
\def\ra{\rightarrow}
\def\Re{{\cal R \mskip-4mu \lower.1ex \hbox{\it e}\,}}
\def\Im{{\cal I \mskip-5mu \lower.1ex \hbox{\it m}\,}}
\def\be{\begin{equation}}
\def\ee{\end{equation}}
\def\ra{\rightarrow}
\begin{titlepage}
\vspace*{-1cm}
\noindent
\phantom{bla}
\hfill{UMHEP-419}
\\
\vskip 2.5cm
\begin{center}
{\Large\bf Two Comparative Case Studies \\
 of $b$-quark and $c$-quark Physics:} \\
\vskip 0.3cm
$\bullet$ {\Large\bf Particle-antiparticle Mixing} \\
$\bullet$ {\Large\bf Radiative Weak Decays
\footnote{talk delivered at XXXth Rencontres de Moriond
(QCD and High Energy Interactions), 19-26 March 1995, Les Arcs,
France} } \\
\end{center}
\vskip 1.6cm
\begin{center}
Eugene Golowich\\
\vskip .1cm
Department of Physics and Astronomy \\
University of Massachusetts, Amherst MA 01003, USA\\
\end{center}
\vskip 2cm
\begin{abstract}
\noindent
We discuss prospects for detecting the two charm-related phenomena of
$D^0$-${\bar D}^0$ mixing and weak radiative decays of $D$ mesons
({\it e.g.} $D\to K^* + \gamma$).  A general update of
particle-antiparticle mixing for the pseudoscalar mesons is presented
and the dynamics of mixing is reviewed, with application especially
to the $D^0$-${\bar D}^0$ system.  The radiative weak decays of
$B$ mesons is then considered, and the problem of hadronic uncertainties
is reviewed.  Finally, the technique of calculating radiative weak
decays for charm mesons is explained.
\end{abstract}
\vfill
\end{titlepage}
\vskip2truecm

Even with the recent discovery of the $t$-quark, we are clearly
in the midst of the `$b$-quark era'.  Two $B$-factories are under
construction and extensive data samples have been produced at
fixed-target accelerators and $e^+ e^-$ colliders.  Although we all
await the day when studies of CP-violation can be carried out,
two discoveries already of special prominence are $B_d$-${\bar B}_d$ mixing
and the $B\to K^* \gamma$ rare decay.  At the same time, however,
there has been an impressive advance in $c$-quark studies.  Yet neither
mixing nor weak radiative decays have been found for $D$ mesons.  We shall
discuss the current status of both phenomena, emphasizing the differences
between $B$-meson and $D$-meson systems.

\begin{center}
{\bf Particle-antiparticle Mixing}
\end{center}

For simplicity, let us ignore the complication of CP-violation.
Then mixing occurs if some interaction can convert a neutral pseudoscalar
meson $P^0$ to its antiparticle ${\bar P}^0$ .  If so, the two
eigenstates $P^{(1)}$ (CP-even) and $P^{(2)}$ (CP-odd) experience
a mass difference $\Delta m$, and oscillations between particle and
antiparticle ensue (with $\omega = \Delta m/2$) from an
initial state of $P^0$ (or ${\bar P}^0$).  Since mesons are
unstable, decay also takes place.  Let us denote the larger decay width
of the two CP eigenstates by $\Gamma$.  The competition between
mixing and decay is characterized by parameter $x \equiv \Delta m
/\Gamma$:
\phantom{xxxx}\vspace{0.07in}
\begin{center}
\begin{tabular}{c||c|c|c|c}
\multicolumn{5}{c}{Table~1 {Current Status of Mixing}$^{\cite{pdg}}$}\\
\hline
Meson & $K^0$  & $B_d$ & $D^0$ & $B_s$ \\ \hline
$x$ & $0.476\ (1\%)$ & $0.71\ (14\%)$ & $< 0.083$ & $ > 2.0$ \\ \hline
\end{tabular}
\end{center}
\phantom{xxxx}\vspace{0.06in}
We see that two mixing parameters are accurately known, but the other two
are only bounded.  To grasp the physical meaning of these numbers, I recommend
plotting $\Im \left[ (\Gamma/2)/(m - E - i\Gamma/2)\right]$ for each CP
eigenstate.  This is simply the line shape as given by the imaginary part of
a Breit-Wigner function.  The graphs accompanying this paper display (for
all the meson systems) the profile of each CP-eigenstate as a function of
energy, with unit of energy chosen to yield a Lorentzian shape of reasonable
width.  Let us comment briefly on each graph:
\begin{enumerate}
\item $K^{(1)}$-$K^{(2)}$: Note the striking difference in widths.
The CP-odd $K^{(2)}$ is narrow due to suppressed phase space.
The spacing between the curves, $\Delta m_K$, is relatively well
understood in terms of the quark box diagram, with some uncertainty associated
with the precise value of the $B_K$ parameter and the role of long-range
effects.  The feature least understood is the width of the broad CP-even
$K^{(1)}$, which is the problem of the $\Delta I = 1/2$ rule.
\item $B_d^{(1)}$-$B_d^{(2)}$: The two curves have almost the same
width since phase space is no longer an issue and there is no dynamical
mechanism in the Standard Model to produce a large $\Delta\Gamma$.
Compared to kaon mixing, all aspects of this diagram are well
understood.
\item $D^{(1)}$-$D^{(2)}$: Although one of the profiles appears to have
been omitted, this is not the case.  In the Standard Model,
$D$-meson mixing is quite feeble and on a plot which
displays the decay width in a natural manner, it is not possible to
separate the $D^{(1)}$ and $D^{(2)}$ peaks!
\item $B_s^{(1)}$-$B_s^{(2)}$: This case is almost opposite
to the above since the mixing oscillation is expected to dominate
decay.  We use equal widths here, although $\Delta\Gamma /\Gamma$
might be in the $0.1\to 0.2$ range.$^{\cite{lopr1}}$
\end{enumerate}
Of course, plotting the particle and antiparticle time dependence as
oscillation/decay occurs is also instructive but limitation of space
prevents us from doing so here.

What is the dynamics of mixing?  Two categories of effects occur,
short-range (quark box-diagrams) and long-range.  Despite uncertainties
in estimating meson decay constants and $B$-parameters, it is believed that
short-range contributions are the most important component of all mixing
amplitudes except perhaps for charm.  As regards $\Delta m_D$, the current
experimental bound and the value of the short-distance component are
respectively
\beq
|\Delta m_D|^{\rm (expt)} < 1.3\times 10^{-10}~{\rm MeV} \quad
{\rm and} \quad |\Delta m_D|^{\rm (s.d.)} \simeq 0.8\times
10^{-14}~{\rm MeV} \ ,
\eeq
implying a gap of about four orders of magnitude!  But there are also
Standard Model long-range effects. Studies of the one-particle ({\it pole})
and some two-particle ({\it dispersive}) intermediate states imply
long-distance values in the range $|\Delta m_D|^{\rm (l.d.)}
\sim 10^{-13}$~MeV.$^{\cite{ld}}$

Yet another possible theoretical approach to $D^0$-${\bar D}^0$ mixing
is application of heavy-quark effective theory.  One expands
in inverse powers of the $c$-quark mass, and generates contributions
$|\Delta m_D|_{\rm HQET}^{(n)}$ for the lowest orders $n = 4,6,8$.
Although there is concern about using HQET at such low energies, we
cite the results (in units of $10^{-14}$~MeV) $|\Delta m_D|_{\rm HQET}^{(n)}
\simeq (0.5\to 0.9)$, $(0.7\to 2.0)$, $(0.1 \to 0.6)$ respectively for
$n = 4,6,8$.$^{\cite{hq}}$  These are smaller than the long-range estimates
just given, and suggest possible cancellations between
the various $n$-particle intermediate states.
\eject
\begin{center}
{\bf Radiative Weak Decays}
\end{center}

Within errors, the recent observations of the exclusive decay $B \to K^*
\gamma$ and the associated inclusive $b\to s \gamma$
transition$^{\cite{cleo}}$ are both in accord with theoretical expectations
(also having error bars!) of the Standard Model.  These findings are important
because of the dominance (due to the large $t$-quark mass) of the loop
({\it penguin}) amplitude.  But there is important work yet to do.  On the
experimental side, the respective uncertainties of $39\%$ and $29\%$ must
be decreased, while theoretically, advances must
occur in computing QCD corrections and in taking the hadronic
matrix element of the QCD-improved penguin operator.

When meaningful comparison between experiment and theory becomes a reality,
the two will either agree or disagree.  Then, either one cites success of
the Standard Model and places limits on models of new physics$^{\cite{new}}$
or one claims observation of new physics.  I wish to caution against
premature acceptance of the latter because the penguin is not
the {\it only} contribution coming from the Standard Model --- there
can be nonspectator contributions as well as those from long
range (nonpenguin) effects.  In Ref.~\cite{gp}, a careful
treatment is given for constructing a gauge-invariant vector-dominance
(VMD) contribution to $B \to K^* \gamma$.  Of course, such
long-range effects are difficult to quantitatively estimate and
one will see a number of increasingly sophisticated calculations
appearing over time.

The theoretical picture of weak radiative decays in charm is quite
different.  The penguin amplitude is tiny, as shown in
Table 2, which lists the intermediate-quark mass dependence of the
Inami-Lim$^{\cite{il}}$  function $F_2$,
first without and then with the CKM dependence.
\phantom{xxxx}\vspace{0.1in}
\begin{center}
\begin{tabular}{ccc}
\multicolumn{3}{c}{Table~2 {Loops in $b\to s \gamma$ and $c\to u\gamma$}}\\
\hline\hline
$b\to s \gamma$ & $F_2$ & $V_{ib}V_{is}^* F_2$ \\ \hline
$u$ & $2.27\times 10^{-9}$ & $1.29\times 10^{-12}$ \\
$c$ & $2.03\times 10^{-4}$ & $7.34\times 10^{-6}$ \\
$t$ & $0.39$ & $1.56\times 10^{-2}$ \\ \hline
$c\to u \gamma$ & $F_2$ & $V_{ci}V_{ui}^* F_2$ \\ \hline
$d$ & $1.57\times 10^{-9}$ & $3.36\times 10^{-10}$ \\
$s$ & $2.92\times 10^{-7}$ & $6.26\times 10^{-8}$ \\
$b$ & $3.31\times 10^{-4}$ & $3.17\times 10^{-8}$ \\ \hline\hline
\end{tabular}
\end{center}
\phantom{xxxx}\vspace{0.1in}
The remarkable influence of the $t$-quark mass in $b\to s \gamma$
is evident, as is also the tiny magnitude of the Inami-Lin function
for $c\to u\gamma$.  A laborious calculation of the QCD-corrected
hamiltonian for $c \to u \gamma$ has been performed in Ref.~\cite{bghp} and
even with a large QCD enhancement, the penguin component to $c\to u\gamma$
is found to be negligible.  Thus, long-range effects will surely dominate
weak radiative decays of charm mesons and the pattern of observed
decays can be expected to provide a stern test of our calculational
abilities.  At present, one has only limits such as
$B_{D^0\to \rho^0\gamma} < 1.4\times 10^{-4}$ and
$B_{D^0\to \phi^0\gamma} < 2.0\times 10^{-4}$.$^{\cite{selen}}$
A large number of radiative weak decays of charm mesons is
addressed in Ref.~\cite{bghp} and the modes $D_s^+\to \rho^0\gamma$,
$D^0\to K^{*0} \gamma$ are especially favored, with branching ratios
in the $10^{-5}\to 10^{-4}$ range.  Unfortunately, these calculations are
even more difficult than for $B$-mesons due to the presence of final-state
interactions in the $c$-quark mass region.

\begin{center}
{\bf Conclusions}
\end{center}

The very fact that neither mixing nor radiative weak decay have yet been
observed in $c$-quark systems make these phenomena inviting targets for
experimental study.  In this talk, we have attempted to provide an
update for each topic.

As regards $D^0$-${\bar D}^0$ mixing, we feel that any experimental
determination of $\Delta m_D$ much larger than $10^{-13}$~MeV would be
grounds for excitement since some kind of new physics would presumably
be responsible.  It is the very suppression in the Standard Model of
weak-mixing for charm that makes this a potentially rewarding area for
study.  However, the level of experimental sensitivity required will be
exceedingly severe!

The situation is brighter for the near-term observation of radiative
weak decays for $D$ mesons, particularly at CLEO with its capability
for detection of photons and neutral mesons.  Rather than placing
bounds on new physics, such findings will undoubtedly be of most interest
in revealing the interplay of weak and QCD effects in the dynamically
interesting charm quark region.  In particular, they could yield significant
insights as to the influence of final-state interactions.$^{\cite{bghp}}$

\end{document}